\documentclass{icrc2009}

\usepackage{graphicx}   
\usepackage{caption}    
\usepackage[font=footnotesize]{subfig} 
\usepackage{fixltx2e}
\usepackage{url}
\usepackage{amsmath,amssymb,amscd}
\usepackage{afterpage}

\newcommand{\shorttitle}[1]%
{\markboth{Proceedings of the 31\MakeLowercase{$^{st}$} ICRC, {\L}\'{o}d\'{z} 2009}{#1} }
\newcommand{\etal}{\MakeLowercase{\textit{et al. }}} 

\begin{document}

\title{Impact of high-energy  hadron interactions \\ on  the  atmospheric
        neutrino flux predictions}

\author{\IEEEauthorblockN{A.~A.~Kochanov\IEEEauthorrefmark{1}\IEEEauthorrefmark{2},
			  T.~S.~Sinegovskaya\IEEEauthorrefmark{1},
        S.~I.~Sinegovsky\IEEEauthorrefmark{1}\IEEEauthorrefmark{2}}                                                                             \\
\IEEEauthorblockA{\IEEEauthorrefmark{1} Institute of Applied Physics, Irkutsk  State  U., Gagarin Bvld. 20, Irkutsk, Russia 664003}
\IEEEauthorblockA{\IEEEauthorrefmark{2}  Depart. of Theor. Physics, Irkutsk  State  U., Gagarin Bvld. 20, Irkutsk, Russia 664003}}

\shorttitle{S.I.Sinegovsky \etal Atmospheric neutrino flux predictions}
\maketitle

\begin{abstract}

We study the influence of hadron interaction features on the high-energy
atmospheric neutrino spectrum. The 1D calculation is performed with use
of the known high-energy hadronic models, SIBYLL 2.1, QGSJET-II, Kimel and Mokhov, for the parameterizations of primary cosmic ray spectra issued from the measurement data.  The results  are compared with the Frejus data and AMANDA-II measurements as well as with other calculations.
Sizable difference of the neutrino fluxes (up to the factor of 1.8 at 1 TeV) that are obtained
with the SIBYLL and QGSJET-II appears to be rather unexpected keeping in mind the hadron and muon flux calculations in the same energy region. 
  
\end{abstract}


\section{Introduction}

\begin{table*}[!th]  
  \caption{Spectrum weighted moments $z_{pc}(E_0)$, calculated for $\gamma=1.7$}
  \label{tab_z}
  \centering \vspace{0.2cm} 
  \begin{tabular}{|c|c||c|c|c|c|c|c|}
  \hline
   Model  & $E_0$, GeV & $z_{pp}$ & $z_{pn}$ & $z_{p\pi^+}$  & $z_{p\pi^-}$  & $z_{pK^+}$  & $z_{pK^-}$  \\
   \hline 
             & $10^2$     & 0.174    & 0.088    & 0.043     & 0.035  & 0.0036   &  0.0030   \\
   QGSJET-II & $10^3$     & 0.198    & 0.094    & 0.036     & 0.029  & 0.0036   &  0.0028   \\
             & $10^4$     & 0.205    & 0.090    & 0.033     & 0.028  & 0.0034   &  0.0027   \\
            \hline
             & $10^2$     & 0.211    & 0.059    & 0.036     & 0.026  & 0.0134   &  0.0014      \\
   SIBYLL 2.1         & $10^3$     & 0.209    & 0.045    & 0.038     & 0.029  & 0.0120   &  0.0022   \\ 
             & $10^4$     & 0.203    & 0.043    & 0.037     & 0.029  & 0.0097   &  0.0026   \\ 
            \hline  
            & $10^2$     & 0.178    & 0.060    & 0.044     & 0.027  & 0.0051  &  0.0015   \\
   KM       & $10^3$     & 0.190    & 0.060    & 0.046     & 0.028  & 0.0052  &  0.0015   \\ 
            & $10^4$     & 0.182    & 0.052    & 0.046     & 0.029  & 0.0052  &  0.0015   \\   
            \hline
  \end{tabular}
  
  \end{table*} 
\begin{figure*}[!t]
   \centerline{\subfloat[Calculation for the GH primary spectrum~\cite{GH}]
{\includegraphics[width=0.39\textwidth, trim = 1.1cm 0.0cm 0.0cm 0.0cm]{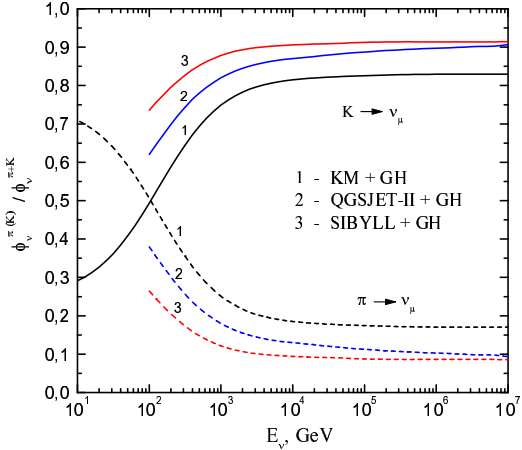}\label{fig-1a}}     
              \hfil
     \subfloat[Calculation for Zatsepin and Sokolskya model~\cite{ZS3C}] 
{\includegraphics[width=0.39\textwidth, trim = 1.1cm 0.0cm 0.0cm 0.0cm]{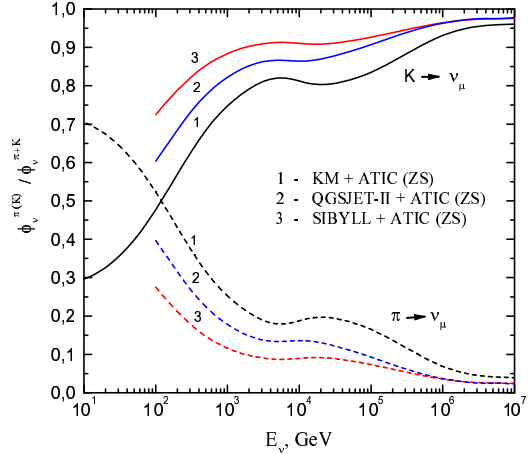}\label{fig-1b}}}  
   \caption{Fraction of $(\nu_{\mu}+\bar\nu_{\mu})$ flux from  kaons (solid lines)  and  pions (dashed) calculated for $\theta=0^\circ$}
    \label{pika_frac}
 \end{figure*}
 
Atmospheric neutrinos (AN) appear in decays of mesons (charged pions, kaons etc.) produced through collisions of  high-energy cosmic rays with air nuclei.  The AN flux in the wide energy range remains the issue of the great interest  since the low energy AN flux is a research matter in the neutrino  oscillations studies, and the high energy atmospheric neutrino flux is now appearing as the unavoidable background for astrophysical neutrino experiments~\cite{nt200_06, nt200_08, amanda05, amanda07, amanda08, amanda09}.
To present day a lot of calculations are made of atmospheric neutrino fluxes, among which~\cite{Volk80, GSB88, BN89, Dedenko89, Lipari93, AGLS96, nss98, FNV2001, Derome03} (see  also~\cite{GHKLMNS96, Naumov2001, GH, BGLRS04, HKKM04} for a review of 1D and 3D calculations of the  AN flux in the wide energy range).

In this work we present results of new one-dimensional calculation of the atmospheric muon neutrino flux in the range $10$--$10^7$ GeV made with use of the hadronic models  QGSJET-II 03~\cite{qgsjet2}, SIBYLL 2.1~\cite{sibyll} as well as  Kimel and  Mokhov (KM)~\cite{KMN} that were tested also in recent atmospheric muon flux calculations~\cite{KSS08}. 
We make an attempt to learn how strongly  the diversity of hadronic interaction models influences on the high-energy spectrum of atmospheric neutrinos.    
 \begin{figure}[!ht]
  \centering
\includegraphics[width=0.37\textwidth, trim = 1.5cm 0.0cm 0.0cm 0.0cm]{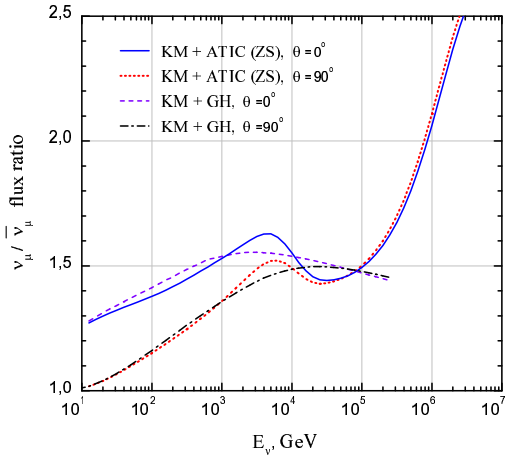} 
  \caption{Ratio of the $\nu_\mu$ and $\bar\nu_\mu$ fluxes calculated with KM model for GH and ZS primary spectra.}
  \label{rat_KM}\vskip -0.5 cm
\end{figure}

\section{The method and input data \label{sec:method}}

 The calculation is performed on the basis of the  method~\cite{NS} of solution of the hadronic cascade equations in the atmosphere,  which takes into account non-scaling behavior of inclusive particle production cross-sections, the rise of total inelastic hadron-nuclei cross-sections, and the non-power law primary spectrum (see also~\cite{KSS08}). 
As the primary cosmic ray spectra and composition in wide energy range used is the model recently proposed  by Zatsepin and Sokolskaya (ZS)~\cite{ZS3C}, which fits well the  ATIC-2 experiment data~\cite{atic2} and supposedly to be valid up to $100$ PeV.
The ZS proton spectrum at $E\gtrsim 10^6$ GeV is compatible with KASCADE data~\cite{KASCADE05} as well the helium  one is  within the range of the KASCADE spectrum recontructed with the help of QGSJET 01 and SIBYLL models.
Alternatively in the energy  range $1-10^6$ GeV we use the parameterization by Gaisser, Honda, Lipari and Stanev  (GH)~\cite{GH, GHLS}, the version with the high fit to the helium data. Note this version is consistent with the data of the  KASCADE experiment at $E_0>10^6$ GeV that was obtained (through the EAS simulations) with the SIBYLL 2.1.

To illustrate the distinction of the hadron models employed in the computations, it is appropriate to compare the spectrum-weighted moments (Table~\ref{tab_z}) computed for proton-air interactions (for $\gamma =1.7$): 
 \begin{equation}\label{mom}
z_{pc}(E_0)
= \int\limits_0^1\frac {x^{\gamma}}{\sigma^{in}_{pA}}\frac {d\sigma_{pc}}{dx}\,dx,
\end{equation}
where  $x=E_c/E_0$, \ $c=p,n,\pi^\pm, K^\pm$.
The values in Table~\ref{tab_z} display approximate scaling law  both  in SIBYLL 2.1 and KM  and little violation of the scaling  in the QGSJET-II for $p$ and $\pi^\pm$.

\section{Atmospheric muon neutrino fluxes\label{sec:an}}

Along with major sources of the muon neutrinos, $\pi_{\mu2}$ and  $K_{\mu2}$ decays, we consider three-particle semileptonic decays, $K^{\pm}_{\mu3}$, $K^{0}_{\mu3}$,  the contribution originated from decay chains   $K\rightarrow\pi\rightarrow\nu_\mu$ ($K^0_S\rightarrow \pi^+\pi^-$, $K^\pm \rightarrow \pi^\pm \pi ^0$), as well as small fraction from the muon decays. 

One can neglect the 3D effects in calculations of the atmospheric muon neutrino flux  
near vertical at energies $E \gtrsim 1$ GeV and  at $E \gtrsim 5$ GeV in case of directions close to horizontal (see~\cite{BGLRS04,HKKM04}).
Fractions of the neutrino flux near vertical from pion and kaon decays 
are shown in Fig.~\ref{pika_frac}. These calculations are made for the model primary spectrum  by GH~\cite{GH} (Fig.~\ref{fig-1a}) as well for the model by ZS~\cite{ZS3C} that comprises  the results of ATIC-2 experiment~\cite{atic2} (Fig.~\ref{fig-1b}). Note the similar ratio for muon fluxes differs from that of neutrino fluxes: at $10^3$ GeV the  $\mu_K/\mu_\pi$  ratio is about 0.25, while the $\nu_K/\nu_\pi$ is about 4 (see also Fig.~4 in Ref.~\cite{GH}).

The ratio $\nu_\mu/\bar\nu_\mu$ calculated with  KM model for the two primary spectra, GH and ZS, is plotted  in Fig.~~\ref{rat_KM}. The wavy shape of the ratios apparently visible in Figs.~\ref{fig-1b} and~\ref{rat_KM} reflects the peculiarities of the ZS spectra. 
  \begin{table}[!h]
  \caption{Ratio of the $\nu_\mu$ fluxes at $\theta=0^\circ$\,($90^\circ$) calculated with the
    SIBYLL 2.1, QGSJET-II, and KM}
  \label{tab_flux}
  \centering\vspace{0.2cm} 
  \begin{tabular}{c||c|c|c} \hline
     $E_{\nu}$, GeV  & $1$ & $ 2 $ & $3$ \\ 
   \hline 
    & \multicolumn{3}{c}{GH} \\ 
    $10^2$  & 1.65 (1.22)  & 0.97 (0.85)  & 1.70 (1.44) \\
    $10^3$  & 1.71 (1.46)  & 0.96 (0.92)  & 1.78 (1.59) \\
    $10^4$  & 1.60 (1.57)  & 0.96 (0.96)  & 1.67 (1.64) \\
    $10^5$  & 1.54 (1.49)  & 0.99 (0.96)  & 1.56 (1.55) \\
  \hline \hline 
      &   \multicolumn{3}{c}{{ZS}}    \\ 
    $10^2$  & 1.58 (1.26)  & 1.00 (0.91) & 1.58 (1.38)\\
    $10^3$  & 1.64 (1.39)  & 0.95 (0.92) & 1.73 (1.51)\\
    $10^4$  & 1.55 (1.46)	 & 0.96 (0.95) & 1.61 (1.54)\\
    $10^5$  & 1.37 (1.23)	 & 0.91 (0.83) & 1.51 (1.48) \\
    $10^6$  & 1.10 (0.95)	 & 0.61 (0.55) & 1.80 (1.73)\\
    $10^7$  & 0.89 (0.75)	 & 0.48 (0.43) & 1.85 (1.74)\\ \hline 
  \end{tabular}
  \end{table}
\begin{figure}[!th]            %
  \centering
\includegraphics[width=0.47\textwidth, trim = 1.1cm 0.0cm 0.0cm 0.0cm]{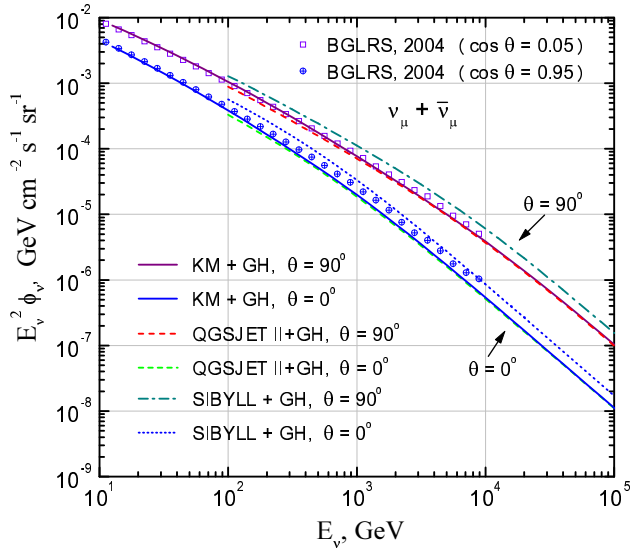}
 \caption{Comparison of the two independent calculations for the GH spectrum.} 
  \label{3mod_GH} \vskip -0.5 cm
 \end{figure}
 \begin{figure}[!b]
  \centering 
 \includegraphics[width=0.47\textwidth, trim = 1.1cm 0.0cm 0.0cm 0.0cm]{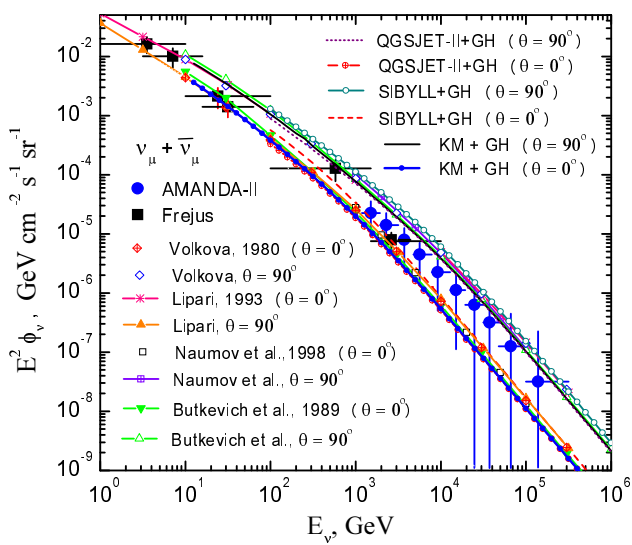} 
\caption{Comparison of the present calculation as well as the previous ones (by Volkova~\cite{Volk80},  Butkevich et al.~\cite{Dedenko89}, Lipari~\cite{Lipari93}, Naumov et al.~\cite{nss98}) with the data of the AMANDA-II~\cite{amanda07} and  Frejus~\cite{frejus} experiments. This work calculation  codes are in the right top corner.}
  \label{all_comp}
 \end{figure}

A comparison of  ($\nu_\mu+\bar\nu_\mu$) flux calculations for the three hadronic models under study is made in     
 Table~\ref{tab_flux}: column 1, 2 and 3  presents the flux ratio, $\phi_{\nu_{\mu}}^{(\rm SIBYLL)}/\phi_{\nu_\mu}^{(\rm KM)}$,  $\phi_{\nu_{\mu}}^{(\rm QGSJET{\text -}II)}/\phi_{\nu_\mu}^{(\rm KM)}$ and  $\phi_{\nu_{\mu}}^{(\rm SIBYLL)}/\phi_{\nu_{\mu}}^{\rm(QGSJET{\text -}II)}$ correspondingly, calculated at $\theta=0^\circ$ and $90^\circ$ (in brackets) with usage of the GH and ZS primary spectrum.  
One can see that usage of QGSJET-II and SIBYLL models leads to apparent difference of the muon neutrino flux, as well as in the case of SIBYLL as compared to KM  (unlike the muon flux, where SIBYLL and KM lead to very similar results ~\cite{KSS08}). 
On the contrary, the QGSJET-II neutrino flux is very close to the KM one: up to $100$ TeV the difference does not exceed $5\%$ for the GH spectrum and $10\%$  for the  ZS one at $\theta=0^\circ$. 
While the muon flux discrepancy in the QGSJET-II and KM predictions is about $30\%$ at vertical~\cite{KSS08}. The origin of differences is evident:  the kaon production ambiguity.  
\begin{figure*}[!th]
  \centering 
\includegraphics[width=0.53\textwidth, trim = 1.1cm 0.0cm 0.0cm 0.0cm]{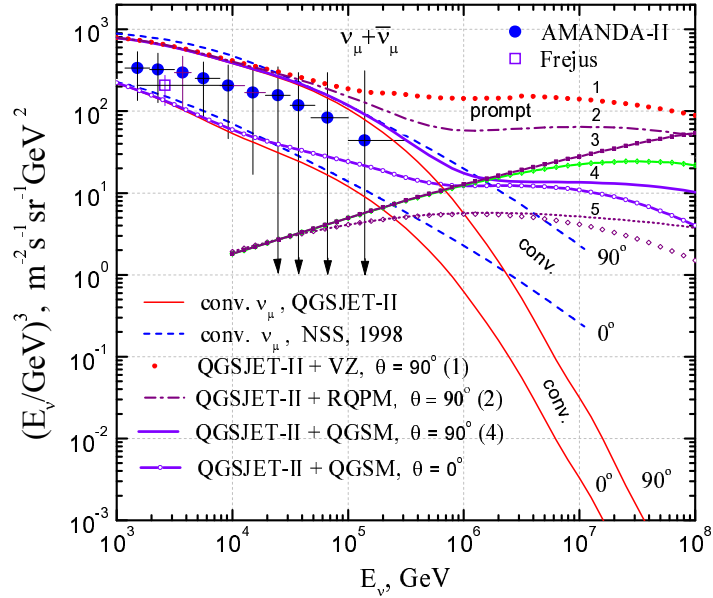} 
  \caption{Fluxes of the conventional and prompt muon neutrinos along with data points from the  AMANDA-II~\cite{amanda07} and Frejus~\cite{frejus} experiments. Codes of curves marking the prompt neutrino flux at $\theta = 90^\circ $ are folows:   1 -- VZ~\cite{VZ01};   2 -- RQPM~\cite{bnsz89}; 
3 -- GGV~\cite{GGV00} (the case of $\lambda=0.5$, where $\lambda$ is exponent of the gluon distribution at low Bjorken $x$) ;  
4 -- QGSM~\cite{bnsz89};   5 -- GGV ($\lambda=0.1$).
 Curves just below the 3, 4 and 5 display the coresponding flux at $\theta = 0^\circ$.}
  \label{pr_nu_5m} \vskip -0.5 cm
 \end{figure*} 
 
 Figure~\ref{3mod_GH} shows this work calculations of the neutrino flux (lines) in comparison with the result of  Barr, Gaisser, Lipari, Robbins and Stanev (BGLRS)~\cite{BGLRS04}) obtained with use of the  TARGET 2.1 (symbols).  All these computations are performed for the GH primary spectra.  One can see the calculations for KM and TARGET 2.1 are in close agreement in the range $10-10^4$ GeV (near horizontal) as well as at $E_\nu < 200$ GeV  near vertical. 
 
In Fig.~\ref{all_comp} presented is the comparison of different calculations of the  AN flux along with the data of the  AMANDA-II~\cite{amanda05, amanda07} and  Frejus~\cite{frejus} experiments. More comparisons of the low and high energy flux calculations may be found in Refs.~\cite{AGLS96,  nss98, GHKLMNS96, Naumov2001, BGLRS04}. 
  
Figure~\ref{pr_nu_5m} presents  the comparison of this work calculation of the conventional (from $\mu, \,\pi,\, K$-decays) and prompt muon neutrino flux with some of previous ones~\cite{nss98, Naumov2001, bnsz89,  VZ01, GGV00}. 
The conventional flux here was calculated with use of QGSJET-II model combined with the Zatsepin and Sokolskaya primary spectrum (thin lines). Dashed lines mark the calculation by Naumov, Sinegovskaya and Sinegovsky~\cite{nss98, Naumov2001} of the conventional muon neutrino fluxes for $\theta = 0^\circ$ and $90^\circ$. 
Bold dotted line (curve 1) shows the sum of the prompt neutrino flux by Volkova and Zatsepin (VZ)~\cite{VZ01} and the conventional one due-to the QGSJET-II + ZS model  at $\theta = 90^\circ$.  Dash-dotted line (2) marks the sum of the  QGSJET-II  conventional  flux ($\theta = 90^\circ$) and the prompt neutrino contribution due to the recombination quark-parton model (RQPM)~\cite{bnsz89}. Solid line 4 shows the same for the prompt neutrino flux due to the quark-gluon string model (QGSM)~\cite{bnsz89} (see also~\cite{nss98, Naumov2001, prd98}).  Also shown are the two of the prompt neutrino predictions by Gelmini, Gondolo and Varieschi (GGV)~\cite{GGV00} (curves 3 and 5).

Notice that recent evaluation of the prompt neutrino flux obtained with the dipole model (DM)~\cite{DM}, is rather close to the QGSM prediction at $E_\nu \gtrsim 10^6$ GeV, keeping in mind that the theoretical uncertainty absorbs a difference of the DM and QGSM fluxes.  

The prompt neutrino fluxes at $E_{\nu}=100$ TeV are presented in Table~\ref{tab_3} along with the upper limit on the  astrophysical muon neutrino diffuse flux obtained in AMANDA-II experiment~\cite{amanda07}.  
Note that the QGSJET-II+GH flux appears to be the most low flux of the conventional atmosperic neutrinos at high energies. 
 \vskip -0.2 cm 
 \begin{table}[!h]
  \caption{Atmospheric neutrino flux at $E_\nu = 100$ TeV vs. the AMANDA-II restriction for the $\nu_\mu+\bar\nu_\mu$ flux}    
  \label{tab_3}
  \centering \vspace{0.2cm} 
  \begin{tabular}{|l|c|} \hline
     Model & $E_\nu^2\phi_\nu$, (cm$^2$\,s\,sr)$^{-1}$ GeV \\ 
   \hline 
   conventional $\nu_\mu+\bar\nu_\mu$ : &   $0^\circ$  \quad\quad\quad       $90^\circ$ \\
  QGSJET-II + ZS  & $1.20\times 10^{-8}$ \quad $10.5\times 10^{-8}$ \\   
  QGSJET-II + GH  & $1.11\times 10^{-8}$ \quad $9.89\times 10^{-8}$   \\ \hline  
   prompt $\nu_\mu+\bar\nu_\mu$ : &  $90^\circ$  \\
   VZ~\cite{VZ01}           & $8.12 \times 10^{-8}$  \\
  RQPM~\cite{bnsz89}        & $4.61 \times 10^{-8}$  \\
 QGSM~\cite{bnsz89}         & $1.22 \times 10^{-8}$    \\
  \hline\hline 
 AMANDA-II upper limit~\cite{amanda07}  & $7.4 \times 10^{-8}$   \\
  \hline  
   \end{tabular}  \vskip -0.2 cm
  \end{table}    

\section{Summary} 
 The calculations of the high-energy atmospheric muon neutrino flux demonstrate rather weak dependence on the primary specrtum models in the energy range $10-10^5$ GeV.  However the picture seems to be less steady because of sizable flux differences originated from  the models of high-energy hadronic interactions. As it can be seen by the example QGSJET-II and SIBYLL 2.1, the  major factor of the discrepancy is the kaon production in nucleon-nucleus collisions.    

A common hope that atmospheric muon fluxes might be reliable tool to promote the discrimination between the hadronic interaction models seems to be rather illusive as the key differences in the $\pi$/$K$ production impact variously on the neutrino flux and muon one. For the high-energy neutrino production at the atmosphere the kaon yield in nucleon-nucleus interactions is more strong factor in comparison with that for production of the atmospheric muons, despite on their common to neutrinos origin. 

Inasmuch as the atmospheric prompt neutrino flux weakly depends on the zenith angle (near $100$ TeV),  one may refer the AMANDA-II restriction just to the prompt neutrino flux model. Thus one may consider both RQPM and QGSM to be consistent 
with the AMANDA-II upper limit for diffuse neutrino flux. \vskip 0.1 cm

We thank V.~A.~Naumov for helpful discussions. We are grateful to T.~Pierog for clarifying comments concerning the codes of hadronic interaction models.  This research was supported in part by the Russian Federation Ministry of Education and Science within the Programme "Development of Scientific Potential in Higher Schools" under grants 2.2.1.1/1483, 2.1.1/1539, Federal Programme "Scientific and educational cadres of innovation Russia", grant P1242,  and by President Programme ``Russian Scientific Schools", grant NSh-1027.2008.2.

\end{document}